# Are There Really Cooper Pairs And Persistent Currents in Aromatic Molecules?

by

R. H. Squire [δ]
N. H. March [κ]
A. Rubio [α]

[δ] Department of Natural Sciences
West Virginia University - Institute of Technology
Montgomery, WV 25136, USA
email: richard.squire@mail.wvu.edu

[κ] Department of Physics
University of Antwerp
Groenborgerlaan 171, B-2020
Antwerp, Belgium
and
Oxford University
Oxford, England

[α] NanoBio Spectroscopy Group and ETSF Scientific Development Centre and DIPC
Department of Materials Science, Faculty of Chemistry
University of the Basque Country  UPV/EHU
Centro Joxe Mari Korta, Avenida de Tolosa, 72,
and
E-20018 Donostia-San Sebastian, Spain
Fritz-Haber-Institut der Max-Planck-Gesellschaft
Faradayweg 4-6  D-14195 Berlin-Dahlem / Germany

**Abstract**

Over twenty years ago one of us suggested the title was affirmative.  In 2012 Cooper pairs were identified in several, but not all "aromatic" compounds tested, benzene being one.  This manuscript discusses the formation of three time-reversed pairs of states forming pseudobosons (high energy Cooper pairs)  in benzene at room temperature. The large stabilization in energy that results is the additive effect of energy gaps of an s wave state and a charge density wave permitting the pseudobosons to exist at room temperature.  The overall result of these interactions is three pseudobosons occupying the lowest boson state and the positions of the carbon  nuclei are optimum by forming a perfect hexagon.  The possibility of a persistent current exists; detection might not be easy.

**1. Introduction.** Over twenty years ago we proposed that the formation of highly correlated electron pairs, possibly a high energy form of Cooper pairs, might explain why certain organic compounds such as benzene have a much more stable ground state than might be expected **by eV** [1, 2]. We were not completely satisfied with our article since the prevalent thought on the upper temperature limit with a phonon mechanism is $T_c \sim 30K$ [3]; today it is speculated to be as high as 170K [4]. Also, our use of perturbation theory seemed inappropriate for such a large energy. Indeed, we were unaware that in the 1950's Feynman tried to deduce a theory of superconductivity by an infinite order perturbation theory summation on the Fröhlich electron-phonon interaction; the sum was zero. He did not succeed because the energy gap related to the electron-phonon interaction has an essential singularity which cannot be solved by a power series [5]. The purpose of this manuscript is to propose a more complete theory of benzene and other organics including recent developments which encompass the gist of the original publication [1]. An outline of this manuscript begins with section 2, a brief review of superconductivity followed by new considerations in section 3. Section 4 discusses topological aspects while section 5 contains a summary, conclusions, and possibilities for future work.

**2. Brief Review.** The original BCS theory [6] (for a recent topical review, see [7])] begins with a rather weak attraction between phonons and electrons described by the Fröhlich Hamiltonian which leads to an instability of the "normal" electron bonding pattern near the boundary between filled energy levels ("Fermi sea") and unfilled levels. The resulting overall energy follows from a competition between the usually dominant Coulomb potential and the weak attraction mentioned. Pines used a crude approximation called the "jellium model" to describe this condition where the solid is approximated by point ions in a fluid of electrons [8],

$$V(\vec{q},\omega) = \frac{4\pi e^2}{q^2 + k_s^2} + \frac{4\pi e^2}{q^2 + k_s^2} \frac{\omega_{\vec{q}}^2}{\omega^2 - \omega_{\vec{q}}^2} \qquad (1)$$

The first rhs term is the Coulomb repulsion (screened by $k_s^2$) and the second term is the phonon induced attraction (second fraction) which is usually negligible. However, if $\omega_{\vec{q}} > \omega$, $V(\vec{q},\omega) < 0$ and attraction can prevail. This illustration is not particularly quantitative (see [9] for quantitative details), but Cooper discovered that any attraction would destabilize the Fermi sea. Further, he found that no matter what size of an interaction matrix he used, there was always a pair of electrons that was separated from the continuum [10]. Schrieffer then discovered how to describe a coherent wave function for these electron pairs which obeyed the Pauli principle [11]. When the transition temperature $T_c$ is reached, the electron pairs (pseudobosons) form and immediately condense into a Bose-Einstein Condensate (BEC) where the ground state is populated by multiple pairs. All of the pairs are highly correlated and have the same phase leading to London's speculation that the wave function is macroscopic in nature [12] which was later defined more precisely by Penrose and Onsager [13, 14]. London also suggested that the macroscopic wave function has a phase rigidity, analogous to the barrier of rotation of a carbon double bond, which is described today by the energy functional[15, 16]:

$$E[\varphi] = \frac{1}{2}\rho_s \int \nabla^2 \varphi \qquad (2)$$

Here $\rho_s$ is the phase stiffness and $\varphi$ is the phase of the macroscopic wave function. Since the phase $\varphi$ is fixed, this results in "symmetry breaking" of the electromagnetic field where the phase is random $\left(e^{i\varphi}, 0 \leq \varphi < 2\pi\right)$, and the uncertainty principle, stated as $\Delta\varphi\Delta N \sim 1$, leads to a variable particle number. To continue with the BCS theory, Bardeen had worked out the boundaries a successful theory of superconductivity should have; he also suggested that the competition (eq 1) could be approximated by a particle in a box expression which greatly simplified the calculations of properties which were incredibly accurate considering nothing approaching a suitable theory has existed for fifty years. There were some discrepancies with the original theory such as the appearance of a lack of gauge invariance, but these issues were soon resolved [17].

**3a. New Considerations. Experimental.** Recent studies have found that pseudobosons are present in only certain "aromatic" compounds such as benzene, naphthalene, anthracene and coronene to name a few, but not in other supposedly "aromatic" compounds [18]. These compounds were investigated after experimental studies on other large molecules [19, 20]. The evidence is a hump in the double to single photoionization ratio at an energy of about 38.3 eV. To obtain this match requires a particle of mass 2e, a pseudoboson; then, the de Broglie wavelength equals an average carbon - carbon bond length of 1.4 A in order to obtain consistency. Thus, it seems appropriate to present some new theoretical considerations in this area and discuss their connections with this and other possible experiments.

**3b. Theoretical.** The benzene correlation energy is estimated to be $4/3$ eV [21, 22, 23], which suggests vibrational modes may be involved in the stability. Coherent phonons have been suggested for some time as correlation partners with electronic modes [24]. Indeed, a charge density wave (CDW) has almost identical equations to a SC since they have the same Peierls' origin [25], so a suitable vibrational correlation partner could be characterized as a CDW which optimizes the nuclei's position to further reduce energy. Considerable literature has been devoted to the competition between a CDW and SC, but several recent Hubbard model studies have suggested that there are regions where they can possibly coexist [26, 27, 28]; other studies are not supportive [29].

Our approach rearrangements deviates from the usual Hückel's solution which is two opposite spin electrons in the same angular momentum state. We choose pairing a +n $\alpha$ spin with –n $\beta$ spin, and vise-versa for all of these orbitals; then, these are pairs of time-reversed states, all with angular moment $l = 0$, resulting in s- (or extended s-) wave states [30] which have an energy gap. This arrangement leads to three pseudobosons which should equilibrate and all occupy the bosonic ground state, the lowest electronic energy state of the system.

If we use second quantization, we can express a s state wave function (eq. 3) as a combination of

$$|\psi_0\rangle = \sum_{k>k_\sigma} g_{\vec{k}} c^*_{\vec{k}\uparrow} c^*_{-\vec{k}\downarrow} |\sigma\rangle \qquad (3)$$

pi orbitals and sigma orbitals $|\sigma\rangle$, and $g_{\vec{k}}$ is a weighting coefficient. The time-reversed pairs of states are manifest and always occupied together, or completely empty. The pi portion in this form can be reduced to a "Cooper pair" orbital wave function [31], the form similar to standard BCS treatments,

$$\psi_{pi}(\vec{r}_1, \vec{r}_2) = \sum_{\vec{k}} g_{\vec{k}} e^{i\vec{k}\cdot\vec{r}} e^{-i\vec{k}\cdot\vec{r}} \Rightarrow \psi_{pi}(\vec{r}_1 - \vec{r}_2) = \left[\sum_{k>k_\sigma} g_{\vec{k}} \cos \vec{k}\cdot(\vec{r}_1 - \vec{r}_2)\right](\alpha_1\beta_2 - \beta_1\alpha_2) \quad (4)$$

A CDW, first proposed by Peierls [32] is a standing wave created by combining electron states moving in opposite directions. The result is a periodic distortion of the lattice (usually 1D or quasi-2D) and the electronic charge density would be unstable to the formation of an energy gap, and the electronic states would be reduced compared to their Fermi energy, $E_F$, similar to superconductivity. Much later, Fröhlich proposed that strong interactions between electrons and phonons could lead to an energy gap at $\pm k_F$ if a phonon has a wavevector equal to $2k_F$ [33]. Then, the electron density is no longer connected to the lattice but is linked to the resonant lattice displacements. He further proposed that a circular CDW system based on the periodic boundary conditions should exist. At stabilization energies the correlated electrons and lattice displacements could move through the lattice in an organized manner carrying a dissipation-free current. This object may have been experimentally observed [34]. Since benzene is a ring, the angular translation symmetry replaces the linear translational symmetry and satisfies the multi-valued Born-von Karman boundary conditions; the angular momentum quantum number can also be viewed as a topological number. A 1D CDW can be expressed as [35]

$$\rho(x) = \rho_0 \cos(2k_F x + \varphi) \quad (5)$$

where $\rho_0$ is the uniform density, $k_F$ is the Fermi wave vector and $\varphi$ is the phase angle. Both the s (or extended s-) wave SC state and the CDW state contain energy lowering gaps which will be strongest when the arguments of the cosines in eqs (4) and (5) adjust to a self-consistent solution so they "lock" in the most energy and the gaps become additive. The slower phonon mode oscillation frequency can be followed by the much faster electron density. The electron correlation can be maximized by a lattice vibration mode if the frequency is near the Debye resonance frequency. The result is that a commensurate CDW forms a perfect hexagon which reinforces the electron correlation between the three bosons in the symmetric ground state. At a resonance condition the lattice charge density could also "overshoot" the response to the Coulomb repulsion (first term in (1)), and further enhance pairing. The standard spectroscopic measurements report their measurements on individual particles created by the uncertainty principle $\Delta\varphi\Delta N \sim 1$ and not the cooperative effects.

The "attraction" needed to bind the pi electrons into pairs can be viewed as reduction of the repulsive electrostatic force. Bardeen's suggests that the formation of bosons is due to the exclusion principle as pairing is the best use of the available phase space to form a low energy, coherent ground state [36]. Because of this type of ground state, benzene should have characteristics of a localized room temperature BEC with strong wave function rigidity and a microscopic condensation in angular momentum space $(l=0)$. The common momentum of the paired states gives the long-range order in momentum space (ODLRO) London proposed; it is just not very macroscopic since it is constrained to the angular momentum space with a coherence length $\xi_0 \sim 8\,\text{Å}$.

A comment is in order about the energy experimentally reported, 38.3 eV. This is a quite large energy relative to the empirical resonance energy possibly due to the number of other sources of collective and local stabilizations which are very difficult to estimate and also the residual nuclear-nuclear energies (see eq 23, [57]). There seems no doubt that pseudobosons are qualitatively measured in these compounds. Two independent measure of their existence is proposed below.

**4. Persistent Currents and Topological Considerations.** Hückel's original solution used cylindrical coordinates $(r, z, \varphi)$ to approximate regular n-polygons so the potential for each electron was a periodic function of $\varphi$ with period $2\pi/n$,

$$V(r,z,\varphi) = V(r,z,\varphi + 2\pi/n \, g) \text{ with } g = 1, 2, ..., n-1, \quad (6)$$

Using Bloch's method (the product of a freely propagating wave multiplied by a spatially periodic function $u^k(r,z,\varphi)$) a spatially periodic function can be generated in the standard way

$$\gamma^k(r,z,\varphi) = e^{ik\varphi} u^k(r,z,\varphi)) \text{ with } k = 0, \pm 1, \pm 2, ... \quad (7)$$

Either of these conditions (or the Born-von Karman boundary condition) will make configuration space non-simply connected, for example as g in eq(6) changes, the potential and the wave function become multi-valued. Since the symmetric ground state for benzene should have a singularity (perhaps weak) in the middle of the molecule, if the molecule is oriented in a strong magnetic field, the flux through the singularity should be quantized in units of $\Phi_0 = hc/e^*$ through the center singularity or "hole" ($e^*$ is the charge transported). While benzene may not be the optimal candidate, certainly coronene would be a better candidate as it also has a spectroscopic Cooper pair and an area about four times as large. If an experiment such as this could be conducted and had enough sensitivity, it would be surprising if it would not show flux quantization in a magnetic field as both superconductors and circular charge density waves do. For both superconductors and sliding CDW's, $e^* = 2e$ [38][39], and [40], respectively.

Perhaps a more attainable experiment would be to orient rigid, dilute planes of benzene ( other larger molecules verified as containing pseudobosons in layers on an appropriate surface with the plane of the molecule parallel with the surface and subject the sample to a magnetic field. A current flow might be generated that would be persistent after removal of the magnetic field since the room temperature charged pseudobosons could be accelerated by a magnetic field. Since there is no dissipation, the bosons could be accelerated until some process interrupted their circulation. If no persistent current is detected at room temperature, the temperature could be lowered and current flow could be measured. Experimental details might be similar to the measurement od persistent currents in metals [41]. Thus type of experiment is **not** a typical solution NMR experiment where random orientation and Brownian motion dissipate the current.

**5. Summary, Conclusions and Possibilities for future work.**

We have presented a new model for a room temperature BEC in benzene and other select "aromatics". The ring structure is crucial as the circular structure forces several energetically favorable conditions which might be otherwise competing in a linear model. The notion of a circular Cooper pair is opposite the normal prescription of oscillating electrons. Perhaps as the energy of a Cooper pair becomes larger at the BCS - BEC crossover, a circular configuration can be more stable as zero-momentum, opposite spin pairs. It would be an interesting result if the quantization of flux or a persistent current could be established in an aromatic that has a spectroscopically observed Cooper pair.